\documentclass{ws-procs975x65}            

\begin{document}
\title{Proper orthogonal decomposition vs. Fourier analysis  for extraction of large-scale structures of thermal convection}

\author{Supriyo Paul}

\address{Computational Fluid Dynamics Team, Centre for Development of Advanced Computing Pune,\\
Pune 411007, India\footnote{Present Address: Department of Physics, A.C. College, Jalpaiguri 735101, India. E-mail: supriyopaul@gmail.com}\\
E-mail: supriyop@cdac.in}

\author{Mahendra K. Verma}

\address{Department of Physics, IIT  Kanpur, Kanpur 208016, India}

\begin{abstract}
We performed a comparative study of extraction of large-scale flow structures  in Rayleigh B\'enard convection  using  proper orthogonal decomposition (POD) and {\em Fourier analysis}. We show that the free-slip basis functions capture the flow profiles successfully for the no-slip boundary conditions.  We observe that the large-scale POD modes capture a larger fraction of total energy than the Fourier modes.  However, the Fourier modes capture the rarer flow structures like flow reversals better.  The flow profiles of the dominant POD and Fourier modes are quite similar.   Our results show that the Fourier analysis provides an attractive alternative to POD analysis for  capturing large-scale flow structures.
\end{abstract}


\bodymatter

\section{Introduction}\label{sec:intro}
 In fluid flows, the large-scale structures  play a major role in its dynamics.  Hence,  an identification of such structures is critical for understanding fluid flows.  Proper Orthogonal Decomposition (POD)~\cite{kosambi,lumley_1967, berklooz_ARFM_1993} is one of the popular methods for this task.  
 In this paper we employ a POD scheme  called ``snapshot method"~\cite{sirovich_1987abc, sirovich_convection}, in which a number of uncorrelated and discrete time snapshots of the flow field are used.

An alternate tool for identifying large-scale structures is  Fourier analysis~\cite{lesiuer:book}, which is relatively easier to compute. Recently, Chandra and Verma~\cite{chandra_PRE_PRL} showed that the Fourier modes play a critical role in the reversal dynamics of turbulent convection. Note that a large number of low-dimensional models have been constructed using the large-scale Fourier modes.  There are several low-dimensional models based on POD modes as well~\cite{podvin_sergent,schumacher}, but  Fourier modes are more amenable for this purpose. 

In the present work, we perform a comparative study between POD and Fourier analysis by employing them to Rayleigh-B\'{e}nard convection in a 2D box.  We will emphasise  similarities and dissimilarities between these two diagnostics tools.

\section{Large-scale Structures of RBC in a Two-dimensional Box}\label{sec:rbc}

We simulate {\em Rayleigh-B\'enard convection} (RBC) in a square box of unit dimension. For the velocity field we assume no-slip boundary conditions ($u = v =0$) on all the walls, and for the temperature field we consider the top and bottom walls to be perfectly conducting, while the side-walls to be insulating. The relevant nondimensionalized equations under Boussinesq approximation for the flow are
\begin{eqnarray}
\partial_t{\bf v} + ({\bf v}\cdot\nabla){\bf v} &=& -\nabla P + RaPrT\hat{y} + Pr\nabla^2{\bf v}, \label{rbc_eqn1}\\
\partial_t T + ({\bf v}\cdot\nabla) T &=& \nabla^2 T, \label{rbc_eqn2}\\
\nabla\cdot{\bf v} &=& 0. \label{rbc_eqn3}
\end{eqnarray}
where ${\mathbf v} = u \hat{x} + v \hat{y}$ is the velocity field, $T$ is the temperature field, the Rayleigh number $Ra$ is the ratio of the buoyancy term and the nonlinear term,  the Prandtl number $Pr$ is the ratio of the kinematic viscosity and the thermal diffusivity, and $\hat{y}$ is the buoyancy direction.

We solve the equations in a unit square box  for Prandtl number $Pr = 1$, and Rayleigh number $Ra = 2\times 10^7$.   For this parameter, the flow is turbulent.  The flow also exhibits flow reversals, that is, probes near the vertical walls exhibit random reversals of the velocity field.  For the simulation we use spectral element code NEK5000~\cite{NEK}, and employ $28 \times 28$ spectral elements with seventh-order polynomials inside each  element.  Thus the overall resolution of the simulation is $196\times 196$ grids.  The aforementioned resolution is sufficient to resolve the boundary layers. We remark that a two-dimensional RBC in an experiment can be realised when the depth of the setup is much smaller compared to the height and the width;   in such systems, the modes along the depth are not generated in a significant manner.

We ran our simulation till the system attains a steady state. During the steady state, we focus our attention on the flow during an interval  from thermal diffusive time $t_A = 12.765$ to $t_B = 12.792$, between which a flow reversal is observed.  We analyse the dominant flow structures during this interval.

\begin{figure}[h!]
\centering
\includegraphics[scale=0.85]{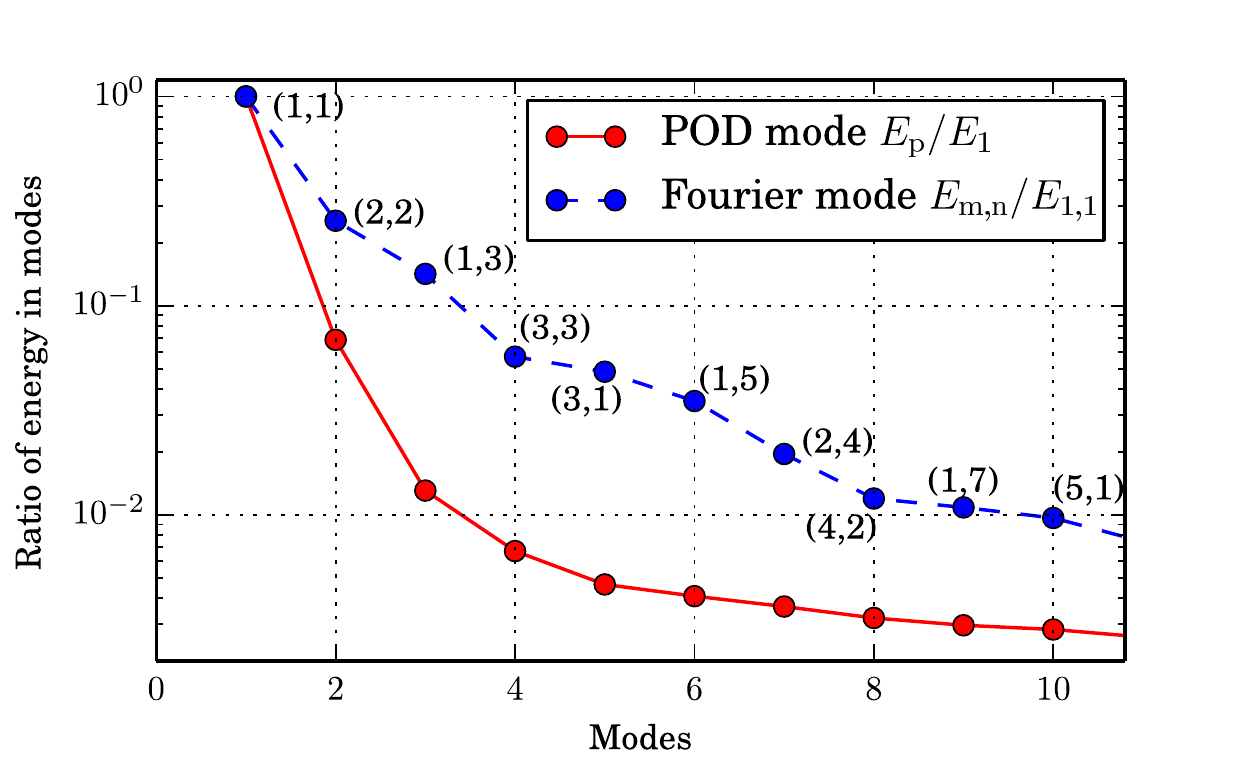}
\caption{Ratio of the energies of different POD modes with the 1st POD mode. Also  the ratio $E_{m,n}/E_{1,1}$, where $E_{m,n}$ is the energy of the $(m,n)$ Fourier mode. }
\label{fig-energy-RBC}
\end{figure}

\begin{figure}[h!]
\centering
\includegraphics[scale=0.75]{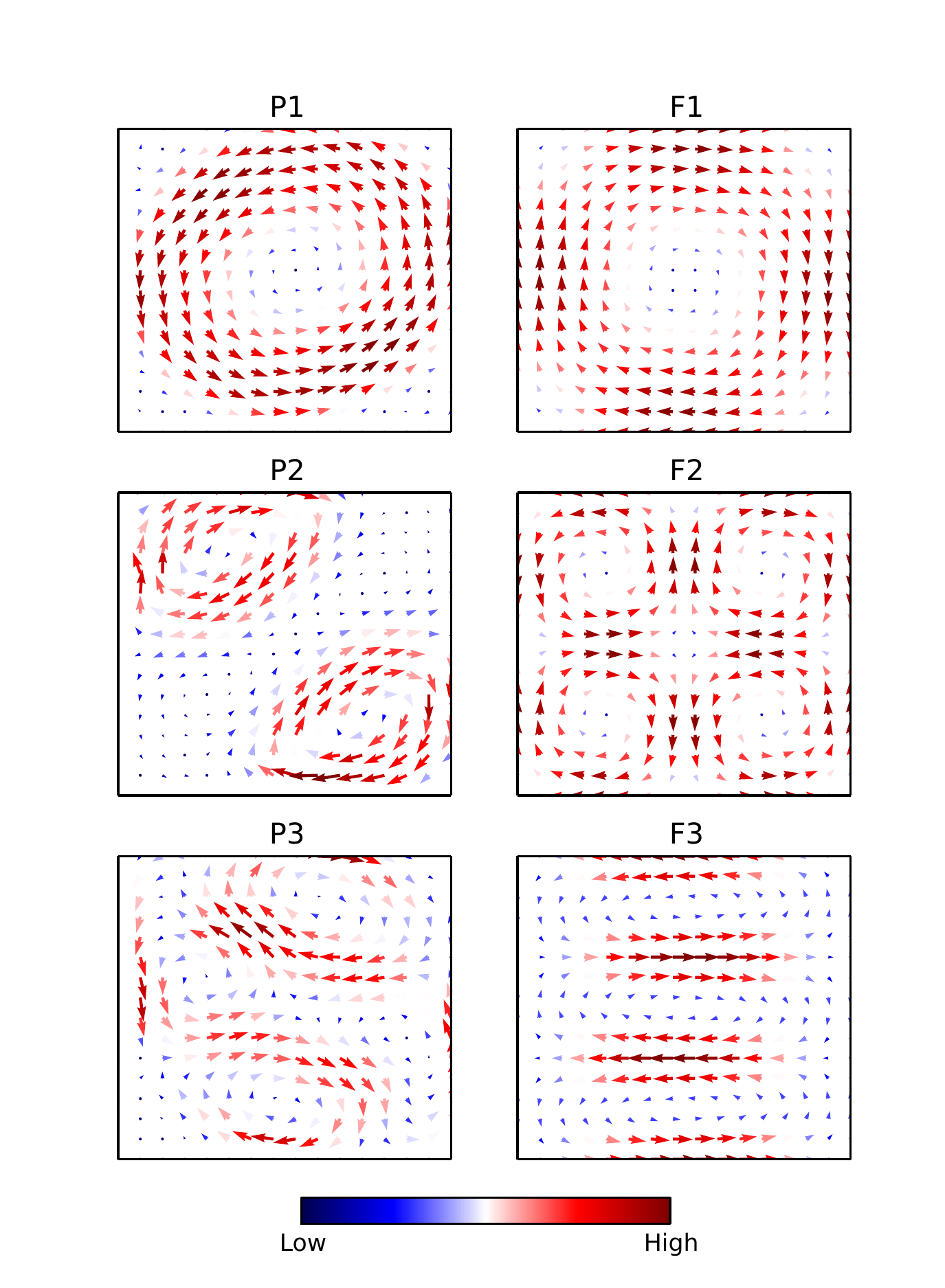}
\caption{Plots of the  three most energetic POD modes (left column) and the corresponding Fourier modes (right column). The top three POD modes are (P1) POD-1, (P2) POD-2, and (P3) POD-3. The top three energetic Fourier modes are (F1) (1,1) mode, (F2) (2,2) mode and (F3) (1,3) mode.}
\label{fig:POD_Four_modes1}
\end{figure}

For the POD analysis\cite{sirovich_1987abc, sirovich_convection} we take 1000 snapshots of the flow in the aforementioned interval in an equidistant manner.  We interpolate the simulation data on a uniform mesh of resolution ($192\times 192$).    For the POD analysis, we construct 1000 vectors using the two components of the velocity and the temperature field.  After this we construct a correlation matrix of these vectors\cite{sirovich_1987abc, sirovich_convection}. We construct the first ten most energetic POD modes using the correlation matrix\cite{sirovich_1987abc, sirovich_convection}.  We extract the first 10 POD modes, which contain 97.5\% of the total energy.  In Fig.~\ref{fig-energy-RBC} we illustrate the energy ratio $E_p/E_1$, where $E_p$ represents the energy of the $p$-th POD mode.   Clearly,  $E_p/E_1$ decreases sharply with $p$ ($E_1/E_2 \sim 14$). Also, the first POD mode contains 88\% of the total energy.  The first three POD modes are exhibited in Fig.~\ref{fig:POD_Four_modes1} as $P1$-$P3$. 

The flow velocity at all the four walls is zero, i.e., ${\bf v} = 0$.  A spectral decomposition of the no-slip boundary condition involves Chebyshev polynomials that resolve the boundary layers quite efficiently.  Note however that the flow structures in the boundary layers are of ``small scales", and they can be ignored while computing the large-scale flow structures.  Chandra and Verma~\cite{chandra_PRE_PRL} observed that the free-slip basis functions defined below
\begin{eqnarray}
u  &=& \sum\limits_{m,n}\hat{u}_{m,n}({\bf k}) \sin(m\pi x) \cos(n\pi y)  \nonumber \\
v  &=& \sum\limits_{m,n}\hat{v}_{m,n}({\bf k}) \cos(m\pi x) \sin(n\pi y)   
\label{eq:free_slip_basis}
\end{eqnarray}
capture the large-scale flow structures quite well. Here $k_x=m\pi$, $k_y=n\pi$ with $m, n$ as positive integers, $\hat{u}_{m,n}, \hat{v}_{m,n}$ are the Fourier components of $u$ and $v$ respectively~\cite{chandra_PRE_PRL}. Note that the free-slip basis functions do not satisfy the no-slip boundary condition.  The success of the free-slip basis functions in capturing the large-scale structures is due to the fact that ignored modes at the boundary layers belong to the small-scale structures.

The temperature field satisfies the conducting boundary conditions at the horizontal plates ($T$=constant), and insulating boundary conditions ($\partial T/\partial x =0$) at the vertical walls.  It is customary to separate $T$ into conducting and convective part, i.e.,
\begin{equation}
T(x,y) = \bar{T}(y) + \theta(x,y)
\end{equation}
where $\bar{T}(y) = 1-y$ is the conduction profile for the nondimensionalized system, and $\theta$ is the temperature fluctuations over $\bar{T}(y) $.  The aforementioned thermal boundary conditions yields:
\begin{eqnarray}  
\theta & = & \sum\limits_{m,n}\hat{\theta}_{m,n}({\bf k}) \cos(m\pi x) \sin(n\pi y)  
\label{eq:T_basis}
\end{eqnarray}

The Fourier transforms can be performed for each snapshot independently, which is one of the main advantages of this analysis, contrary to POD that requires many snapshots.  However to study the dynamics and evolution of the flow structures during a reversal, we study 1000 snapshots of the flow.  We compute the Fourier modes for each frame [Eqs.~(\ref{eq:free_slip_basis}, \ref{eq:T_basis})]. The energy of the higher wavenumber modes decreases sharply,  consistent with the Kolmogorov theory of turbulence.    Since our focus is on the large-scale structures, we take the first 10 Fourier modes that contain approximately  93\% of the total kinetic energy. In Fig.~\ref{fig-energy-RBC} we plot the ratio $E_{m,n}/E_{1,1}$, where $E_{m,n}$ represents the kinetic energy of the $(m,n)$ Fourier mode.  The Fourier analysis reveals that the modes $(1,1)$, $(2,2)$, and $(1,3)$ are the most dominant modes in the flow.  The first Fourier mode  has 58\% of the total energy.    The Fourier modes of the flow are shown in right column  of Fig.~\ref{fig:POD_Four_modes1} as $F1$-$F3$.

When we compare the two methods for the extraction of large-scale flow structures, we observe that  POD modes are more optimal than the Fourier modes.  For example, the first 10 POD and first 10 Fourier modes  contain respectively 97.5\% and 93\%  of the total energies.  Also, the first POD mode contains 88\% of the total energy, but the corresponding Fourier mode contains  only 58\% of the total energy.  However, the flow structures of the first three POD modes are distinctly similar  {in their shape to the first three Fourier modes, but the higher order modes differ.  Note however that the directions of the velocity fields of the POD and Fourier modes are  anti-correlated, hence the amplitudes of these modes are also opposite to each other (to be discussed below).  

 The amplitude of a POD mode is computed by projecting  the snapshot to the POD mode.  The time series of the first five Fourier and POD modes are exhibited in Fig.~\ref{fig-timeseries-Four-POD} that shows that the amplitudes of the POD and Fourier modes are anti-correlated to each other. This is consistent with the anti-correlation of the velocity fields for the corresponding POD and Fourier modes (see Fig.~\ref{fig:POD_Four_modes1}).  During the flow reversal between  $t=12.78$ and $t=12.785$, the first Fourier mode, as well as the first POD mode, change sign.  The second and third modes also show sharp variations during the flow reversal~\cite{chandra_PRE_PRL}.  The fourth and fifth Fourier modes also show noticeable variations during the reversal process, but the corresponding POD modes do not show noticeable variations during this period. This is due to the fact that top three POD modes are most dominant in the flow, while the fourth and the fifth POD modes are quite weak. Thus both Fourier and POD modes provide valuable information on the dynamics of flow reversal, details of which can be found in  Sergent and Podvin~\cite{podvin_sergent,schumacher}, and  Chandra and Verma~\cite{chandra_PRE_PRL}.

The reconstruction of the snapshots using the POD and Fourier modes is shown in Fig.~\ref{fig-recons-Four-POD} where we show the 5-th and 670-th snapshots of the flow.  The 5-th snapshot is reproduced accurately by both the POD and Fourier  analysis.  However for the 670-th snapshot, which depicts the corner-roll flow structure, POD reconstruction is poorer than the Fourier one. The reason for the discrepancy is due to the averaging  process of the POD analysis. On the average, the first POD mode is more dominant than the second POD-mode ($E_2/E_1 = 1/14$). For the 670-th snapshot, the second POD mode is the most important mode, but its contribution towards the reconstruction of the snapshot get suppressed by the most dominant first POD mode. On the other hand, Fourier modes correctly reconstructs the 670-th snapshot since each snapshot has its own set of Fourier modes, and the (2,2) mode is more dominant than the (1,1) mode for the 670-th snapshot. Hence Fourier reconstruction is better than the POD reconstruction for dynamic flows.

\begin{figure}[h!]
\centering
\includegraphics[scale=0.85]{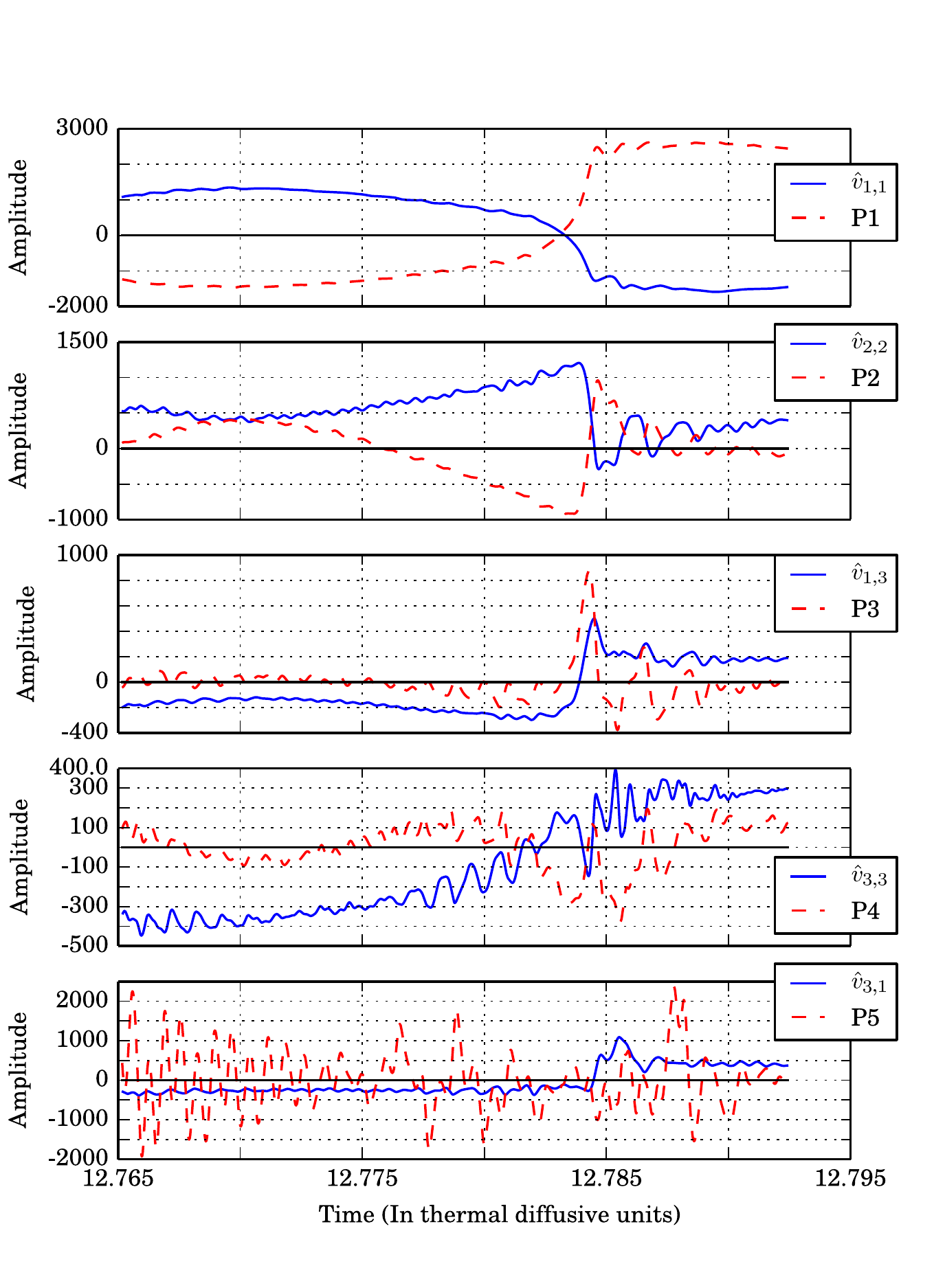}
\caption{ Plot of the time series of the most energetic five  Fourier and POD modes of the vertical velocity $v$ during the reversal. POD coefficients are scaled appropriately for visual clarity.}
\label{fig-timeseries-Four-POD}
\end{figure}

\begin{figure}[h!]
\centering
\includegraphics[scale=0.95]{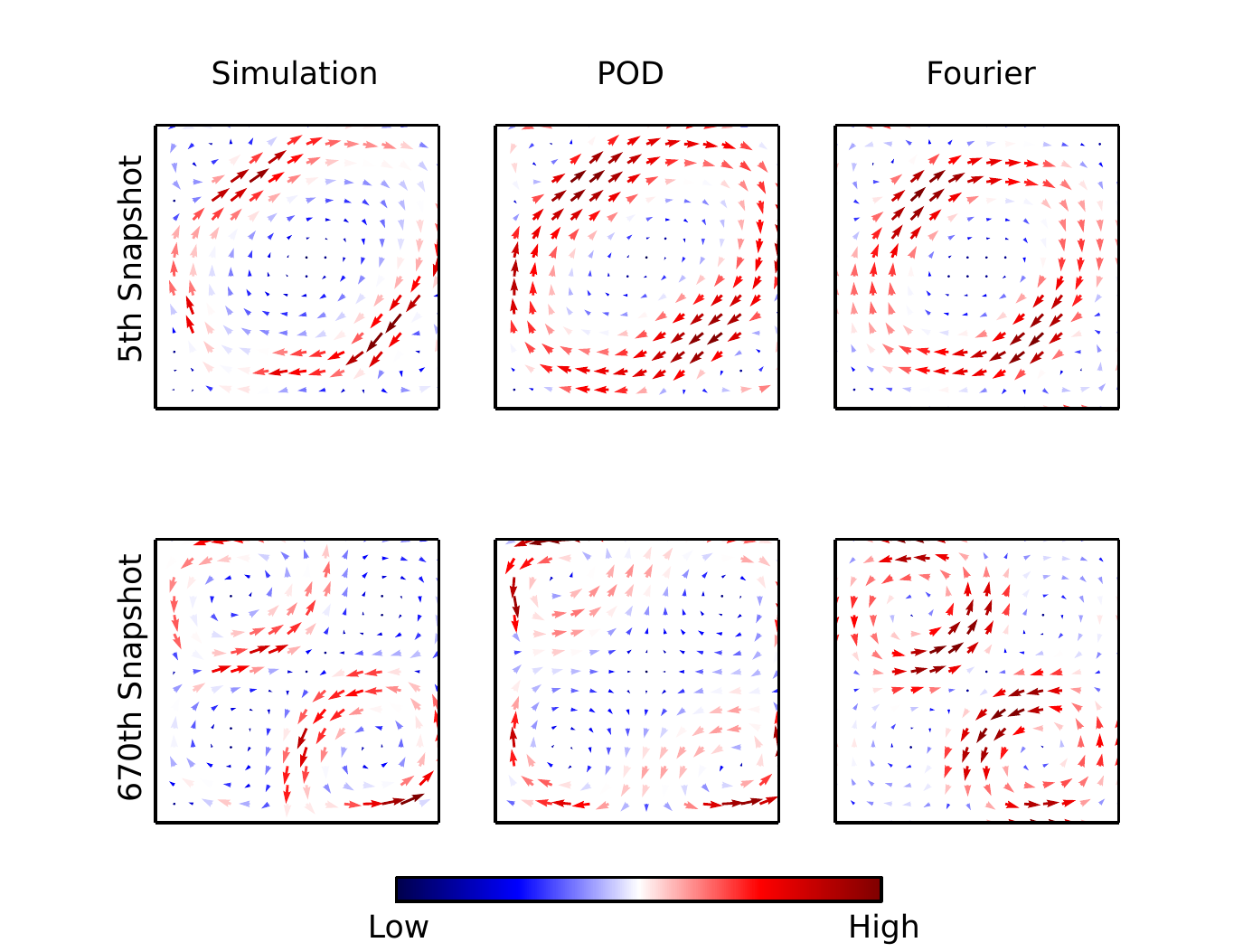}
\caption{Reconstruction of the 5th and 670th snapshots using the first ten POD and ten Fourier modes. Both the POD and Fourier reconstruction are reasonably good, but the the 670th snapshot is better reconstructed using the Fourier modes.}
\label{fig-recons-Four-POD}
\end{figure}

\section{Conclusions and Discussions}\label{sec:conclusion}
In this paper we proposed that the low-wavenumber Fourier modes are good candidates for identifying large-scale flow structures, and they provide an alternative to POD analysis.  In support of our argument, we perform a comparative study of the POD and Fourier analysis of 1000 snapshots of the flow profiles in a two-dimensional Rayleigh B\'{e}nard convection. 

The low wavenumber Fourier modes capture the large-scale structures of the flow quite well.  However,  the first $N$ POD modes contain more energy than the corresponding Fourier modes. Due to this, a reproduction of a flow pattern requires fewer  POD modes than the Fourier modes, but the difference  is not very significant.   Note however that Fourier modes can be computed for each snapshot separately, but the POD analysis requires a large set of snapshots.  A simple estimation shows that the Fourier analysis is computationally less expensive than the POD analysis.  This is useful while dealing with large three-dimensional datasets. Moreover, the Fourier modes have simpler visual interpretations than the POD modes.  

A major objection to the usage of Fourier analysis is its inapplicability to no-slip boundary conditions.  However, the small-scale structures of a boundary layer do not contribute significantly to the large-scale structures. The RBC example discussed in this paper and in Chandra and Verma~\cite{chandra_PRE_PRL} show that free-slip basis can capture the large scale flow structures even under no-slip geometries quite well.

Thus, POD and Fourier analysis have their own advantages and disadvantages.  For idealised geometries like a box, Fourier analysis provides an attractive  alternative to POD.

\section*{Acknowledgment}

The authors thank M. Chandra for sharing with us the RBC data.  SP thanks C-DAC's CFD group  for support and encouragement. The analysis was performed on PARAM-YUVA.


\begin{thebibliography}{99}

\bibitem{kosambi} D. D. Kosambi,  {\it J. Ind. Math. Soc.}, {\bf 7}, 76 (1943).

\bibitem{lumley_1967} J. L. Lumley, ``The structure of inhomogeneous turbulent flows,'' in Atmospheric Turbulence and Radio Wave Propagation, (Nauka, Moscow), p. 166, (1967).

\bibitem{berklooz_ARFM_1993} G. Berkooz, P. Holmes, and J. L. Lumley, {\it Annu. Rev. Fluid Mech.} {\bf 25}, 539 (1993).

\bibitem {sirovich_1987abc} L. Sirovich, {\it Q. Appl. Math.} {\bf 45}, 561 (1987); L. Sirovich, {\it Q. Appl. Math.} {\bf 45}, 573 (1987); L. Sirovich, {\it Q. Appl. Math.} {\bf 45}, 583 (1987).

\bibitem {sirovich_convection} L. Sirovich, L., Maxey, and H. Tarman, ``Analysis of turbulent thermal  convection'',  {\it In Sixth Symposium on Turbulent Shear Flow}, Toulouse, France, p. 68. Springer (1989); A. E. Deane and L. Sirovich, {\it J. Fluid Mech.} {\bf 222}, 231 (1991).

\bibitem{lesiuer:book}  M. Lesieur,``{Turbulence in Fluids}",  (Kluwer Academic, Dordrecht), (1990).

\bibitem{chandra_PRE_PRL} M. Chandra and M. K. Verma, {\it Phys. Rev. E} {\bf 83}, 067303 (2011); M. Chandra and M. K. Verma, {\it Phys. Rev. Lett.} {\bf 110}, 114503 (2013)



\bibitem{podvin_sergent} B. Podvin and A. Sergent, {\it Phys. Fluids} {\bf 24}, 105106 (2012); {\it J. Fluid Mech.} {\bf 766}, 172 (2015).

\bibitem{schumacher} J. Bailon-Cuba, O. Shishkina, C. Wagner, and J. Schumacher, {\it Phys. Fluids} {\bf 24}, 107101 (2012).

\bibitem{NEK} Paul F. Fischer, James W. Lottes, and Stefan G. Kerkemeier, ``{nek5000} {W}eb  page", http://nek5000.mcs.anl.gov, (2008).

\end{thebibliography}
\end{document}